\documentstyle[epsf,12pt]{article}
\newlength{\mytopmargin}
\newlength{\myleftmargin}
\renewcommand{\r}{{\vec{r}\,}}
\newcommand{\F}{{\vec{F}}}
\newcommand{\Q}{{\cal Q}}

\makeatletter
\@addtoreset{equation}{section} 
\makeatother

\begin{document}

\def\thefootnote{}

\begin{center}
{\Large \bf Exact Finite Size Study of the \\[.3cm]
2dOCP at $\Gamma = 4$ and $\Gamma = 6$}
\footnote{\normalsize LPENS-Th 07/99}
\end{center}
\vspace{5mm}

\setcounter{footnote}{0}
\def\thefootnote{\arabic{footnote}}
\noindent
G.~T\'ellez\footnote{email: gtellez@physique.ens-lyon.fr}\\
\noindent
Ecole Normale Sup\'{e}rieure de Lyon, Laboratoire de Physique,
\\Unit\'e Mixte de Recherche 5672
du Centre National
de la Recherche Scientifique, \\
46, all\'{e}e d'Italie, 69364 Lyon Cedex 07, France
\vspace{5mm}

\noindent
P.J.Forrester\footnote{email: matpjf@maths.mu.oz.au}\\
\noindent
Department of Mathematics and Statistics, 
University of Melbourne, \\Parkville, Victoria
3052, Australia

{\small
\begin{quote}
An exact numerical study is undertaken into the finite $N$ calculation
of the free energy and distribution functions for the 
two-dimensional one-component plasma. Both disk and sphere geometries
are considered, with the coupling $\Gamma$ set equal to 4 and 6. 
Extrapolation of our data for the free energy is consistent with
the existence of a universal term ${\chi \over 12} \log N$, where $\chi$
denotes the Euler characteristic of the surface, as predicted theoretically.
The exact finite $N$ density profile is shown to give poor agreement
with the contact theorem relating the density at contact and potential
drop to the pressure in the thermodynamic limit. This is understood
theoretically via a known finite $N$ version of the contact theorem.
Furthermore, the ideas behind the derivation of the latter result are
extended to give a sum rule for the second moment of the pair correlation
in the finite disk, which in the thermodynamic limit converges to the
Stillinger-Lovett result.
\end{quote}
}

\section{Introduction}
The {two-dimensional} {one-component} plasma (2dOCP) is a model in
classical statistical mechanics which consists of $N$ mobile point
particles of charge $q$ interacting on a surface with uniform
neutralizing background charge density. The pair potential
$\Phi(\vec{r},\vec{r}\,')$ between particles is the solution of the
Poisson equation on the particular surface. In the plane
\begin{equation}\label{1.1}
\Phi(\vec{r},\vec{r}\,') = -\log\Big (|\vec{r} - \vec{r}\,'|/l \Big),
\label{1.1ab}
\end{equation}
where $l$ is some arbitrary length
scale which will henceforth be set to unity. With the potential (\ref{1.1})
and a uniform background of charge density $-\rho_b$ inside a disk of
radius $R$ $(\rho_b = N/\pi R^2)$ the corresponding Boltzmann factor,
which consists of the particle-particle, particle-background and
background-background interaction, is given by
\begin{equation}\label{1.2}
e^{-\Gamma N^2  ( (1/2)\log R - 3/8  )}
e^{- \pi \Gamma \rho_b \sum_{j=1}^N |\vec{r}_j|^2/2}
\prod_{1 \le j <k \le N} |\vec{r}_k - \vec{r}_j|^\Gamma,
\end{equation}
where $\Gamma := q^2/k_BT$ is the coupling. We remark that with $\Gamma/2$
an odd integer, (\ref{1.2}) is proportional to the absolute value
squared of the celebrated Laughlin trial wave function for the fractional
quantum Hall effect \cite{La83}.

At the analytic level our knowledge of the properties of the 2dOCP comes
from two main sources. First, for the special coupling $\Gamma = 2$, the
exact free energy and correlation functions can be calculated for a
number of different geometries \cite{AJ81,Ch81,Ca81,JT98}. 
Second, the 2dOCP is an example
of a Coulomb system in its conductive phase and as such should obey a
number of sum rules (see e.g.~\cite{Ma88}) which typically represent
universal properties of such a system. We remark also that the exact
solutions at $\Gamma = 2$ have been an important source of inspiration to
identify universal properties.

In this paper we develop exact numerical solutions at the special couplings
$\Gamma = 4$ and $\Gamma = 6$ for values of $N$ up to 11 and 9 respectively.
By undertaking this study we are able to test the prediction of
Jancovici et al.~\cite{JMP94} that the expression for the free energy
$F$ as a function of the number of particles $N$ be of the form
\begin{equation}\label{1.F}
\beta F = A N + B N^{1/2} + {\chi \over 12} \log N + \cdots,
\end{equation}
where $\chi$ denotes the Euler characteristic of the surface
($\chi = 1$ for a disk, $\chi = 2$ for a sphere). Furthermore we
are able to investigate the rate of convergence of the one and two point
correlation to their thermodynamic values, as well as the accuracy of certain
sum rules in the finite system. In fact the latter line of investigation leads
us to a new sum rule valid for general $\nu$ dimensional multicomponent
Coulomb systems in a spherical domain, which relates to the second moment
of the density-charge correlation function in the finite system. We
recall (see e.g.~\cite{Ma88}) that in the infinite system the second
moment of the charge-charge correlation function is of a universal form
known as the Stillinger-Lovett condition. Indeed our sum rule
(\ref{eq:thenewsumrule}) below gives the finite size correction to this universal
form in systems with a background.

As an outline of the paper, we note here that in Section 2 formulas
are presented specifying the partition function and one and two point
distribution functions for the disk and sphere geometries, with the
coupling an even integer, in terms of certain expansion coefficients.
These expansion coefficients are in general computationally expensive, 
but reasonably efficient algorithms exist in the literature
applicable to the cases $\Gamma = 4$ and 6. Our numerical results
our presented in Section 3. The new sum rules are derived and discussed
in Section 4, while Section 5 concludes with a summary.

\section{Formalism}\label{sec:formalism}
\setcounter{equation}{0}
Our interest is in the exact numerical computation of the partition function
and one and two-point correlation functions for the 2dOCP in a disk and on
the surface of a sphere. In the former system the Boltzmann factor is
given by (\ref{1.1}). Two versions of this model will be considered: one
in which the particles are confined to a disk of radius $R$ (the same
disk which contains the smeared out neutralizing background), and the
other in which the particles are can move throughout the plane. 
These will be referred to as the hard disk and soft disk respectively.
In the
latter system the Boltzmann factor (\ref{1.1}) is assumed valid also
for $|\vec{r}_i| \ge R$, even though the one body potential
$\pi \rho_b |\vec{r}_i|^2/2$ is not the correct potential for the coupling
between a particle and the background in this region (according to Newton's
theorem outside the disk the background creates the same potential as a
charge $-N$ at the origin, so the correct Coulomb potential outside
the disk is $N \log  |\vec{r}_i|$).

On the surface of the sphere the Boltzmann factor is given by
\begin{equation}\label{2.1}
{\left({1\over{2R}}\right)}^{N\Gamma/2}e^{\Gamma N^2/4}\prod_{1\leq j<k\leq
N}{|u_kv_j-u_jv_k|}^\Gamma,
\end{equation}
where $u:=\cos(\theta/2)e^{i\phi/2},\ v:=-i\sin(\theta/2)e^{-i\phi/2}$ are
the Cayley-Klein parameters and $(\theta,\phi)$ are the usual spherical
coordinates. For our purpose it is convenient to consider the stereographic
projection of this system from the south pole of the sphere to the plane
tangent to the north pole. This is specified by the equation
\begin{equation}\label{2.2}
z = 2R e^{i \phi} \tan {\theta \over 2}, \qquad z = x + i y.
\end{equation}
We then have
\begin{eqnarray}\label{2.3}\lefteqn{
{\left({1\over{2R}}\right)}^{N\Gamma/2}e^{\Gamma N^2/4}\prod_{1\leq j<k\leq
N}{|u_kv_j-u_jv_k|}^\Gamma dS_1 \cdots dS_N
= {\left({1\over{2R}}\right)}^{N\Gamma/2}} \nonumber \\&&
\times e^{\Gamma N^2/4}
\prod_{j=1}^N {1 \over (1 + |z_j|^2/(4R^2))^{2 + \Gamma (N-1)/2}}
\prod_{1\leq j<k\leq
N} \left| \frac{z_j - z_k}{2R} \right|^\Gamma d\vec{r}_1 \cdots d\vec{r}_N.
\end{eqnarray}

\subsection{The cases $\Gamma = 4p$}
For $\Gamma = 4p$, integrals over the Boltzmann factors (\ref{1.1}) and
(\ref{2.3}) can be performed from knowledge of the coefficients in the
expansion
\begin{equation}\label{2.S1}
\prod_{1 \le j < k \le N}(z_k - z_j)^{2p} = \sum_\mu
c_\mu^{(N)}(2p) m_\mu(z_1,\dots,z_N)
\end{equation}
where $\mu = (\mu_1,\dots,\mu_N)$ is a partition of $pN(N-1)$ such that
$$
2p(N-1) \ge \mu_1 \ge \cdots \mu_N \ge 0
$$
and
$$
 m_\mu(z_1,\dots,z_N) = {1 \over \prod_i m_i!}
\sum_{\sigma \in S_N} z_{\sigma(1)}^{\mu_1}\cdots
z_{\sigma(N)}^{\mu_N}
$$
is the corresponding monomial symmetric function (the $m_i$ denote the
frequency of the integer $i$ in the partition). The key point for the utility
of (\ref{2.S1}) is that with $z_j = r_j e^{i \theta_j}$, the $m_\mu$ are
orthogonal with respect to angular integrations:
\begin{eqnarray}\label{15.o1}
\int_0^\infty dr_1 \, r_1 g(r_1^2) \cdots \int_0^\infty dr_N \, r_N g(r_N^2)
\int_0^{2 \pi} d \theta_1 \cdots \int_0^{2 \pi} d\theta_N \,
 m_\mu(z_1,\dots,z_N) \overline{m_\kappa(z_1,\dots,z_N)} \nonumber \\
= \delta_{\mu,\kappa} {N! \over  \prod_i m_i!} \pi^N \prod_{l=1}^N G_{\mu_l} \hspace{2cm}
\end{eqnarray}
where $ G_{\mu_l} := 2 \int_0^\infty dr \, r^{1 + 2\mu_l} g(r^2)$ for
arbitrary $g(r^2)$. Thus, after also noting that
\begin{equation}\label{2.PD}
\prod_{j < k}|z_k - z_j|^{4p} = \prod_{j < k}(z_k - z_j)^{2p}
\prod_{j < k}(\bar{z}_k - \bar{z}_j)^{2p},
\end{equation}
we see that for $\Gamma = 4p$
\begin{equation}\label{2.A*}
I_{N,\Gamma}[g] := \int_{{\mathbf{R}}^2}d\vec{r}_1 \, g(r_1^2) \cdots
 \int_{{\mathbf{R}}^2}d\vec{r}_N \, g(r_N^2) \, \prod_{j < k}
|\vec{r}_k - \vec{r}_j|^{\Gamma} = N! \pi^N \sum_\mu
{(c_\mu^{(N)}(2p))^2 \over \prod_i m_i!}  \prod_{l=1}^N G_{\mu_l}.
\end{equation}
In the case $p=1$ this formalism has been utilized by Samaj et 
al.~\cite{SPK94}, 
who
furthermore presented an algorithm for the computation of $\{c_\mu\}$ in
this case. Let us now consider this latter point.

In general the coefficients $c_\mu^{(N)}(2p)$ can be calculated from the formula
\begin{equation}\label{2.t}
c_\mu^{(N)}(2p) = {1 \over (2 \pi)^N} \int_0^{2 \pi} d\theta_1 \,
e^{-i \mu_1 \theta_1} \cdots \int_0^{2 \pi} d\theta_N e^{-i\mu_N \theta_N}
\prod_{j < k} (e^{i \theta_k} - e^{i \theta_j})^{2p},
\end{equation}
which follows from (\ref{2.S1}). Since we require $|\mu| = pN(N-1)$, the
integral over $\theta_N$ can be performed by changing variables
$\theta_j \mapsto \theta_j + \theta_N$ $(j=1,\dots,N-1)$ to give
\begin{eqnarray}\label{2.tt}
c_\mu^{(N)}(2p) & = & {1 \over (2 \pi)^{N-1}} \int_0^{2 \pi} d\theta_1 \,
e^{-i \mu_1 \theta_1} \cdots \int_0^{2 \pi} 
d\theta_{N-1} e^{-i\mu_{N-1} \theta_{N-1}}
\prod_{j=1}^{N-1}(1 - e^{i\theta_j})^{2p} \nonumber \\ && \times
\prod_{1 \le j < k \le N - 1} (e^{i \theta_k} - e^{i \theta_j})^{2p}.
\end{eqnarray}

The simplest case is $N=2$, when the sum over pairs in (\ref{2.tt}) is not
present. Expanding $(1 - e^{i\theta_1})^{2p}$ according to the binomial
theorem gives
$$
c_\mu^{(2)}(2p) = (-1)^{\mu_1} \left ( {2p \atop \mu_1} \right )
$$
where $\mu_1 = p,p+1,\dots,2p$ (for $\mu_1 = p$ we have $\mu_1 = \mu_2$
and thus $m_{\mu_1} = 2$, while in all other cases $\mu_1 \ne \mu_2$ and so
$m_{\mu_1} = m_{\mu_2} = 1$). Substituting in (\ref{2.A*}) we see, after
some minor manipulation, that
\begin{equation}\label{2.2'}
 \int_{{\mathbf{R}}^2}d\vec{r}_1 \, g(r_1^2) 
 \int_{{\mathbf{R}}^2}d\vec{r}_2 \, g(r_2^2) \, 
|\vec{r}_2 - \vec{r}_1|^{4p}
= \pi^2 \sum_{\mu=0}^{2p} \left ( {2p \atop \mu} \right )^2
\int_0^\infty dr \, r^\mu g(r) \int_0^\infty dr \, r^{2p - \mu} g(r).
\end{equation}
To calculate $c_\mu^{(N)}(2p)$ via this method for a general value of $N$ would
require expanding ${1 \over 2}(N-1)N$ products via the binomial theorem,
giving a total of $({1 \over 2}(N-1)N)^{2p+1}$ terms to determine each
value of $c_\mu$. Thus for a given value of $N$ the complexity increases
exponentially with the coupling $p$. As we want to determine the
$c_\mu$ for a sequence of values of $N$ as large as possible, we are
therefore restricted to the case $p=1$.

In fact
the case $p=1$ allows (\ref{2.t}) to be computed without using the binomial
expansion \cite{SPK94}. Instead one uses the Vandermonde formula for the
product of differences as a determinant  to expand the products in
(\ref{2.t}). This gives
\begin{equation}\label{2.van}
c_\mu^{(N)}(2) = \sum_{P \in S_N} \varepsilon(P) \sum_{Q \in S_N} \varepsilon(Q)
\prod_{k=1}^N \delta_{P(k) + Q(k) - 2,\mu_k} =
\sum_{P \in S_N} \varepsilon(P) \sum_{Q \in S_N}
\prod_{k=1}^N \delta_{P(k) + k - 2,\mu_{Q(k)}},
\end{equation}
which is the formula we used 
to compute our data in the case $p=1$ for $N=3,\dots,10$.

\subsection{The cases $\Gamma = 4p+2$}
With $\Gamma = 4p+2$, decomposing the product of differences analogous
to (\ref{2.PD}) shows that we must consider the product of differences
raised to an odd power. The analogue of (\ref{2.S1}) is then the
expansion
\begin{equation}\label{15.1'}
\prod_{1 \le j < k \le N} (z_k - z_j)^{2p+1} =
\sum_\mu c_\mu^{(N)}(2p+1) {\cal A}(z_1^{\mu_1 + N - 1}z_2^{\mu_2 + N - 2}
\cdots z_N^{\mu_N})
\end{equation}
where $2p(N-1) \ge \mu_1 \ge \mu_2 \ge \cdots \ge \mu_N \ge 0$,
$\sum_{j=1}^N \mu_j = p N (N -1)$ and ${\cal A}$ denotes
antisymmetrization. Factoring out the antisymmetric factor
$\prod_{j < k}(z_k - z_j)$ from both sides then gives
\begin{equation}\label{15.2'}
\prod_{1 \le j < k \le N} (z_k - z_j)^{2p} =
\sum_\mu c_\mu^{(N)}(2p+1) S_\mu(z_1,\dots,z_N)
\end{equation}
where $S_\mu$ denotes the Schur polynomial
indexed by the partition $\mu$. Furthermore, analogous to the
orthogonality (\ref{15.o1}) we have
\begin{eqnarray}\label{15.o2}
\int_0^\infty\cdots\int_0^\infty
\prod_{l=1}^N dr_l\, r_l g(r_l^2)
\int_0^{2 \pi}d\theta_1 \cdots \int_0^{2 \pi} d\theta_N 
\prod_{j<k}\left|z_j-z_k\right|^2
 S_\mu(z_1,\dots,z_N) \overline{S_\kappa(z_1,\dots,z_N)} 
\nonumber \\
= \delta_{\mu,\kappa} 
N!\pi^N \prod_{l=1}^N G_{\mu_l+N-l}. \hspace{2cm}
\end{eqnarray}
Thus for $\Gamma = 4p+2$, instead of (\ref{2.A*}) we have
\begin{equation}\label{2.B*}
I_{N,\Gamma}[g] 
 = N! \pi^N \sum_\mu
(c_\mu^{(N)}(2p+1))^2  \prod_{l=1}^N G_{\mu_l+N-l}.
\end{equation}
According to (\ref{15.1'}) the coefficients $c_\mu^{(N)}(2p+1)$ can be
computed from the formula (\ref{2.t}) with $\mu_j \mapsto \mu_j + N -
j$ and $2p \mapsto 2p+1$, or equivalently (\ref{2.tt}) with the same
replacements. In the case $N=2$ this latter formula gives $$
c_\mu^{(2)}(2p+1) = (-1)^{\mu_1 + 1} \left ( {2p +1 \atop
\mu_1 + 1} \right )
$$
with $\mu_1 = p,\dots, 2p$. This in turn implies that the formula
(\ref{2.2'}) again holds with $2p \mapsto 2p+1$.

To obtain data for consecutive values of $N$, the computationally simplest
case is $p=1$. However algorithms based on 
(\ref{2.t}) (with $\mu_j \mapsto \mu_j + N - j$ and
$2p \mapsto 2p+1$) are inferior to methods that determine $c_\mu^{(N)}(3)$
from (\ref{15.2'}) 
\cite{DGIL94,Du94,STW94}. The most efficient algorithm appears to be
the one of Scharf et al.~\cite{STW94}, where the coefficients $c_\mu^{(N)}(3)$
are determined up to $N=9$. Fortunately the authors of \cite{STW94} have
kindly supplied us with their data (up to $N=8$), so we do not need to
repeat the calculation.

\subsection{The sphere}
The Boltzmann factor for the sphere, stereographically projected onto
the plane, is given by the r.h.s.~of (\ref{2.3}). Thus, with
$\r_j\mapsto2R\r_j$ we require
\begin{equation}\label{2.g1}
g(r^2)  = (1 + r^2)^{-(N-1)\Gamma/2 - 2}
\end{equation}
in the integral (\ref{2.A*}). However, computational savings can be obtained
by first noting that because the sphere is homogeneous, one particle can be
fixed at the north pole, reducing the number of integrals from $N$ to
$N-1$ (we must also multiply by $\pi$ -- the area of the surface of a
sphere of radius $1/2$). Thus we have
\begin{eqnarray}
\int_{({\mathbf{R}}^2)^N} d\vec{r}_1 \cdots d\vec{r}_N \,
\prod_{i=1}^N {1 \over (1 + |z_i|^2)^{(N-1)\Gamma/2 + 2}}
\prod_{1 \le j < k \le N} |z_k - z_j|^\Gamma \qquad \nonumber \\
 = \pi \int_{({\mathbf{R}}^2)^{N-1}} d\vec{r}_1 \cdots d\vec{r}_{N-1} \,
\prod_{i=1}^{N-1}  {|z_i|^\Gamma \over
 (1 + |z_i|^2)^{(N-1)\Gamma/2 + 2}}
\prod_{1 \le j < k \le N-1}  |z_k - z_j|^\Gamma,
\end{eqnarray}
and so should choose
\begin{equation}\label{2.g2}
g(r^2) = {r^\Gamma \over (1 + r^2)^{(N-1)\Gamma / 2 + 2} }
\end{equation}
in  (\ref{2.A*}).

With $g(r^2)$ given by (\ref{2.g2}), the formulas (\ref{2.A*}) and
(\ref{2.B*}) show that at $\Gamma = 4$ and $\Gamma = 6$
the canonical partition function
$$
Z_{N,\Gamma} := {1 \over N!}  \int_{({\mathbf{R}}^2)^{N}}
 d\vec{r}_1 \cdots d\vec{r}_N \, e^{-\beta U}
$$
can be represented by the series
\begin{eqnarray}
Z_{N+1,4}^{\mathrm{sphere}} & = & {e^{(N+1)^2} \pi^{N+1} \over N + 1}
\sum_\nu \left(c_\nu^{(N)}(2)\right)^2 {1 \over \prod_i m_i!}
\prod_{i=1}^N {(\mu_i + 2)!(2N - \mu_i - 2)! \over (2N+1)!} \label{2.s1}\\
Z_{N+1,6}^{\mathrm{sphere}} & = & \rho_b^{(N+1)/2} (N+1)^{(N+3)/2}
e^{3(N+1)^2/2} \pi^{3(N+1)/2}
\sum_\nu  (c_\nu^{(N)}(3))^2 \nonumber \\ && \times \prod_{k=1}^N
{(3+N + \mu_k - k)! (2N - 3 - \mu_k + k)! \over (3N + 1)!}. \label{2.s2}
\end{eqnarray}
To obtain these formulas use has been made of the definite integral
\begin{equation}\label{2.defi}
\int_0^\infty {r^p \over (1 + r)^q} \, dr = 
{\Gamma(p+1) \Gamma (q - p - 1) \over \Gamma (q) }.
\end{equation}

Because the sphere is homogeneous, 
the two-point distribution  $\rho_{(2)}((\theta,\phi),
(\theta',\phi'))$ can be computed with one particle at the north pole
($\theta' = 0$ say). We then have 
$$\rho_{(2)}((\theta,\phi), \,
(\theta',\phi')) = \rho_{(2)}(\theta)
$$ 
so the two-point function can be
computed from an integral of the form (\ref{2.A*}). In fact with
$g(r^2)$ given by (\ref{2.g1}) we have
\begin{equation}
\rho_{(2)}(\theta) = 
\frac{1}{4R^2}
\frac{1}{I_{N,\Gamma}[g]}\lim_{x'\to0}
\frac{g(x^2)g({x'}^2)}{4\pi^2 xx'}
(1+x^2)^2(1+x'^2)^2
\frac{\delta^2 I_{N,\Gamma}[g]}{\delta g(x^2) \delta g(x'^2)}
\end{equation}
where $x=\tan(\theta/2)$. For $\Gamma = 4$ this gives
\begin{eqnarray}
\lefteqn{\rho_{(2)}(\theta)
=\rho_b^2\frac{(2N-1)!}{N^2(1+x^2)^{2N-2}}}
\nonumber\\
&\times&\frac{\displaystyle
\sum_{\mu,\mu_N=0}(c_\mu^{(N)}(2))^2 {1 \over \prod_i m_i!} 
\prod_{i=1}^N \mu_i!(2N-2-\mu_i)!
\sum_{k=1}^{N-1} \frac{x^{2\mu_k}}{\mu_k!(2N-2-\mu_k)!}
}{
\sum_{\mu}(c_\mu^{(N)}(2))^2 {1 \over \prod_i m_i!}
\prod_{i=1}^N \mu_i!(2N-2-\mu_i)!
}
\end{eqnarray}
while for $\Gamma = 6$ we deduce that
\begin{eqnarray}
\lefteqn{\rho_{(2)}(\theta)
=\rho_b^2\frac{(3N-2)!}{N^2(1+x^2)^{3N-3}}
\,(3N-2)
}
\nonumber\\
&\times&\frac{
\displaystyle
\sum_{\mu,\mu_N=0} \!\!(c_\mu^{(N)}(3))^2 \!
\prod_{i=1}^N\!
(\mu_i\!\!+\!\!N\!\!-\!\!i)!(2N\!\!-\!\!3\!\!-\!\!\mu_i\!\!+\!\!i)!
\!\sum_{k=1}^{N-1} \frac{x^{2(\mu_k+N-k)}}%
{(\mu_k\!\!+\!\!N\!\!-\!\!k)!(2N\!\!-\!\!3\!\!-\!\!\mu_k\!\!+\!\!k)!}}{
\sum_{\mu}(c_\mu^{(N)}(3))^2
\prod_{i=1}^N (\mu_i+N-i)!(2N-3-\mu_i+i)!
}
\end{eqnarray}

\subsection{The disk}
In the case of the disk, (\ref{1.2}) with $\vec{r}_j \mapsto
R \vec{r}_j$ shows we require
\begin{equation}\label{2.g2'}
g(r^2) = \chi(r) e^{-\Gamma N |\vec{r}_j|^2/2}
\end{equation}
where $\chi=1$ for $r^2 < 1$ and zero otherwise in the case of the hard disk,
while $\chi=1$ for all $r$ in the case of the soft disk. Thus from
(\ref{2.A*}) we have at $\Gamma = 4$
\begin{eqnarray}
Z_{N,4}^{\mathrm{soft\ disk}}& = &
e^{3N^2/2}\left(\frac{1}{2N}\right)^{N^2}
\pi^N
\sum_\mu (c_\mu^{(N)}(2))^2 \left(\prod_i m_i!\right)^{-1}
\prod_{i=1}^N \mu_i! \label{3.sa}\\
Z_{N,4}^{\mathrm{hard\ disk}}& = &
e^{3N^2/2}\left(\frac{1}{2N}\right)^{N^2}
\pi^N
\sum_\mu (c_\mu^{(N)}(2))^2 {1 \over \prod_i m_i !}
\prod_{i=1}^N \gamma(\mu_i+1,2N) \label{3.sb}
\end{eqnarray}
while at $\Gamma = 6$ use of (\ref{2.B*}) gives
\begin{equation} \label{3.sc}
Z_{N,6}^{\mathrm{hard}} = \rho_b^{N/2}N^{-3N^2/2}3^{-N(3N-1)/2}
\pi^{3N/2}e^{9N^2/4}
\sum_\mu (c_\mu^{(N)}(3))^2 \prod_{k=1}^N
\gamma(\mu_k+N-k+1,3N)
\end{equation}
with the soft disk case  obtained by replacing the incomplete gamma
functions by complete gamma functions.

Unlike the situation with the sphere, the density is a non-constant function
in the disk geometry. Now, with $g(r^2)$ given by (\ref{2.g2}) we have
$$
\rho_{(1)}(r) = {g(r^2) \over 2 \pi r}
{\delta \log Z_{N,4}^{\mathrm{disk}} \over \delta g(r^2)}.
$$
At $\Gamma = 4$ this gives
\begin{eqnarray}\label{2.den}
\lefteqn{
\rho_{(1)}(r)=2\rho_b e^{-2\pi \rho_b r^2}
}&&
\nonumber\\
&&\times
\frac{\displaystyle
\sum_\mu (c_\mu^{(N)}(2))^2 { 1 \over \prod_i m_i!}
\prod_{j=1}^N \gamma(\mu_j+1,2N)
\sum_{k=1}^N \frac{(2\pi \rho_b r^2)^{\mu_k}}%
{\gamma(\mu_k+1,2N)}
}{
\sum_\mu (c_\mu^{(N)}(2))^2 { 1 \over \prod_i m_i!} 
\prod_{j=1}^N \gamma(\mu_j+1,2N)
}
\end{eqnarray}
while at $\Gamma = 6$ one obtains
\begin{eqnarray}
\lefteqn{
\rho_{(1)}(r)=3\rho_b e^{-3\pi \rho_b  r^2}
}&&
\nonumber\\
&&\times
\frac{\displaystyle
\sum_\mu (c_\mu^{(N)}(3))^2
\prod_{j=1}^N \gamma(\mu_j+N-j+1,3N)
\sum_{k=1}^N \frac{(3\pi \rho_b r^2)^{\mu_k+N-k}}%
{\gamma(\mu_k+N-k+1,3N)}
}{
\sum_\mu (c_\mu^{(N)}(3))^2 
\prod_{j=1}^N \gamma(\mu_j+N-j+1,3N)
}
\end{eqnarray}
The corresponding formulas for the soft disk are obtained by replacing the
incomplete gamma functions by complete gamma functions.

Finally, we consider the two-point function in the disk geometry. In general
this quantity is not just a function of the distance between particles,
and so we cannot use the formalism based on the orthogonalities
(\ref{15.o1}) and (\ref{15.o2}). However, with one of the particles fixed
at the origin ($\vec{r}\,' = \vec{0}$ say) we have $\rho_{(2)}(\vec{r},
\vec{r}\,') = \rho_{(2)}(r)$, so in this case the formalism used to
compute the densities can again be used. Thus using the general
formula
$$
 \rho_{(2)}(r) = {1 \over Z_{N,\Gamma}}
\lim_{r' \to 0} {g(r^2) g({r'}^2) \over 4 \pi r r'}
{\delta^2 Z_{N,\Gamma} \over \delta g(r^2) \delta g({r'}^2)},
$$
we find for the hard disk case
\begin{eqnarray}\label{eq:2.rho2-g4}
\lefteqn{
\rho_{(2)}(r) =
4 \rho_b^2 e^{-2\pi \rho_b r^2}
}&&
\nonumber\\
&&\times
\frac{
\sum_{\mu,\mu_N=0}
(c_\mu^{(N)}(2))^2 {1 \over \prod_i m_i !}
\prod_{j=1}^{N-1} \gamma(\mu_j+1,2N)
\sum_{k=1}^{N-1}
\frac{(2\pi \rho_b r^2)^{\mu_k}}{\gamma(\mu_k+1,2N)}
}{
\sum_{\mu}
(c_\mu^{(N)}(2))^2 {1  \over \prod_i m_i !} 
\prod_{j=1}^{N} \gamma(\mu_j+1,2N)
}
\end{eqnarray}
\begin{eqnarray}\label{eq:2.rho2-g6}
\lefteqn{
\rho_{(2)}(r) =
9\rho_b^2 e^{-3\pi \rho_b r^2}
}&&
\nonumber\\
&&\times
\frac{\displaystyle
\sum_{\mu,\mu_N=0}
(c_\mu^{(N)}(3))^2
\prod_{j=1}^{N-1} \gamma(\mu_j+N-j+1,3N)
\sum_{k=1}^{N-1}
\frac{(3\pi \rho_b r^2)^{\mu_k+N-k}}{\gamma(\mu_k+N-k+1,3N)}
}{
\sum_{\mu}
(c_\mu^{(N)}(3))^2
\prod_{j=1}^{N} \gamma(\mu_j+N-j+1,3N)
}
\end{eqnarray}
for $\Gamma = 4$ and $\Gamma = 6$ respectively. Again the corresponding
results for the soft disk are obtained by replacing the incomplete
gamma functions by complete gamma functions.

\section{Numerical results}
\setcounter{equation}{0}
\subsection{Free energy -- sphere geometry}
In the Introduction it was commented that the free energy is expected
to have a large $N$ expansion of the form (\ref{1.F}) with $\chi = 2$
in sphere geometry. In fact the constant $B$ in (\ref{1.F}), which is
a surface free energy, should be identically zero in sphere geometry, so
we expect a large $N$ expansion of the form
\begin{equation}\label{3.F1}
\beta F = A N + {1 \over 6} \log N + C + \cdots.
\end{equation}
As noted by Jancovici et al.~\cite{JMP94}, the validity of (\ref{3.F1}) can
be explicitly demonstrated at $\beta = 2$ because of an exact solution due
to Caillol \cite{Ca81}. The mechanism for the exact solution can 
be seen within the present formalism. Thus, at $\Gamma = 2$ we require the
coefficients $c_\mu^{(N)}(1)$ in (\ref{15.1'}). But this follows
from the Vandermonde expansion (recall (\ref{2.van})), which gives
$c_\mu^{(N)}(1)=1$ for $\mu = 0^N$ and $c_\mu^{(N)}(1) = 0$ otherwise.
Substituting in (\ref{2.B*}) with $g(r^2)$ given by (\ref{2.g2}),
and making use of (\ref{2.defi}) we thus obtain \cite{Ca81}
\begin{equation}\label{3.ca}
Z_{N,2}^{\mathrm{sphere}} = \pi^{-N/2} N^{N/2} \rho_b^{-N/2} e^{N^2/2}
\prod_{k=1}^N {(N-k)! (k-1)! \over N!}.
\end{equation}
This substituted into the general formula
\begin{equation}\label{3.fr}
\beta F_{N, \Gamma} = - \log Z_{N, \Gamma}
\end{equation}
leads to the expansion \cite{JMP94}
\begin{equation}\label{3.fr1}
\beta F = Nf + {1 \over 6} \log N + {1 \over 12} - 2 \zeta'(-1) +
{\mathrm{o}}(1)
\end{equation}
where $f = {1 \over 2} \log (\rho_b / 2 \pi^2)$. We remark that by introducing 
the Barnes $G$ function according to
$$
G(z+1) = \Gamma(z) G(z), \quad G(1) = 1
$$
we can write
$$
\prod_{k=1}^N (k-1)! = G(N + 1).
$$
The large $N$ expansion of the Barnes $G$ function is known to be \cite{Ba00}
\begin{equation}\label{3.barnes}
G(N+1) \sim {N^2 \over 2} \log N - {3 \over 4} N^2 + {N \over 2} \log 2 \pi
- {1 \over 12} \log N + \zeta'(-1) - {1 \over 720 N^2} + {\mathrm{O}}\Big (
{1 \over N^4} \Big ).
\end{equation}
This together with Stirling's formula allows us to extend (\ref{3.fr1})
to the expansion
\begin{equation}\label{3.fr2}
\beta F = Nf + {1 \over 6} \log N + {1 \over 12} - 2 \zeta'(-1) +
{1 \over 180 N^2} +  {\mathrm{O}}\Big ({1 \over N^4} \Big ).
\end{equation}

In the cases $\Gamma = 4$ and $\Gamma = 6$, by following the numerical
procedure detailed in the previous section, we have been able to compute
the partition functions (\ref{2.s1}) and (\ref{2.s2}) up to 11 and 9
particles respectively. The results are listed in Table \ref{t3.1}. 
Our results are presented in decimal form. However the terms in the
summations of (\ref{2.s1}) and (\ref{2.s2}) are all rational numbers,
and we have also calculated the sum itself as a rational number. A point
of interest is the factorization of the denominator and numerator of
the rational number. The exact result (\ref{3.ca}) shows that at $\Gamma =
2$ only small integers occur in this factorization. However our exact
data shows that this feature is no longer true at $\Gamma = 4$ or
$\Gamma = 6$. For example, at $\Gamma = 4$ and with $N=9$ we find that
the summation in (\ref{2.s1}) is given by the ratio of primes
$$
\frac{19\cdot 23\cdot 31\cdot 404431651134013\cdot 56827}%
{2^{28}3^{12}5^7 7^8 11^8 13^8 17^8}.
$$

\begin{table}
\caption{\label{t3.1} Exact numerical computation of the expressions
 (\ref{2.s1}) and (\ref{2.s2}) (in the latter case we have set $\rho_b =1$),
and the corresponding free energy (\ref{3.fr}).
}

\vspace{.4cm}
{\small
\begin{tabular}{l|l|l}
$N$ & $Z_{N,4}$ & $\beta F_{N,4}$ \\
\hline
   3 &  9.770695753081390794542103296367E+02 &
-6.884557862719257767291929292830 \\
   4 &  1.081868103379375397165672403770E+04 &
-9.289029644211538110263324038604 \\
   5 &  1.209528877878741526102013133936E+05 &
-11.70315639163470461293716934684 \\
   6 &  1.360835037494310939624360869217E+06 &
-14.12360906745006986750189927991 \\
   7 &  1.537846289459171693753614603094E+07 &
-16.54847857521316551691816164401 \\
   8 &  1.743564157878398325393942744018E+08 &
-18.97661212873318180330363390257 \\
   9 &  1.981770773388678655915061613417E+09 &
-21.40725661197234419004446417460 \\
  10 &  2.257011016434890100740949944465E+10 &
-23.83989230877186989649422160272 \\
  11 &  2.574639922522006241714385546434E+11 &
-26.27414571135846506646694529338 \\
\end{tabular}
}
 
\vspace{.4cm}
{\small
\begin{tabular}{l|l|l}
$N$  &          $Z_{N,6}$ &                                 $\beta F_{N,6}$ \\ 
\hline
 2&  781.80154948970530457541038293910180&  -6.661600935308419284761353568226471 \\ 
 3&  24731.016946702464115291740435512837& -10.115813481655518642906626162676076
\\
 4& 798906.45662411908447403801186279894& -13.590999142330226359670889161470696
\\
 5& 25990836.664099377843271224794515169& -17.073254597869416657276355484106596
\\
 6& 851167572.30792422833993160492670601&  -20.562119579383207945093207167461793\\
 7&  27989023411.960800446597844273994987& -24.055078249259894430456119939885817\\

 8& 923260788226.64381072982338145761830& -27.551177575665397081224942401207047
\\
 9& 30529687045074.352434196537904510620& -31.049720671888250916196597607309575
\\
\end{tabular}
}
\end{table}

To analyze our data we first sought fitting sets of consecutive values of $N$
to the ansatz
\begin{equation}\label{3.ans}
\beta F_{\Gamma, N} = A_\Gamma N + K_\Gamma \log N + C_\Gamma.
\end{equation}
The results are contained in Table \ref{t3.2}. Notice that at $\Gamma = 4$
the value of the free energy per particle $A$ appears to have converged
to 3 decimal place accuracy, while the value of $K$ appears similarly to
be converging, with the final value in the table differing from $1/6$
only in the  third decimal. The general trends are the same for the
$\Gamma = 6$ data, although the convergence rate (as determined by the 
difference between sequential values) is slower.

\begin{table}
\caption{\label{t3.2} Fitting the values of $\beta  F_{\Gamma, N}$ with
$N$ as specified, taken from Table \ref{t3.1}, to the ansatz (\ref{3.ans}).
}

\vspace{.4cm}
\begin{tabular}{l|l|l|lr|l|l|l}
$N$ & $A_4$ & $K_4$ & $C_4$ & \qquad & $A_6$ & $K_6$ & $C_6$ \\ \hline
3,4,5&  -2.447509 & 0.149600 & 0.293616& & 
 -3.526411 & 0.178065  & 0.267797 \\ 
4,5,6&  -2.448705& 0.154963 & 0.290968&&
 -3.506699 & 0.109543 & 0.283938 \\
5,6,7& -2.449038& 0.156787& 0.289696& &
-3.515359& 0.145316 & 0.269664 \\
6,7,8&  -2.449271 & 0.158300 & 0.288384& &
-3.516438 & 0.152316 & 0.263596 \\
7,8,9& -2.449423 & 0.159440 & 0.287231 & &
 -3.516820 & 0.155176 & 0.260704 \\
8,9,10 &  -2.449524 & 0.160290 & 0.286264 & &
& & \\
9,10,11 &  -2.449594 & 0.160960 & 0.285428 & &
& & \\
\end{tabular}
\end{table}

Next we sought fitting four consecutive values of $N$ to the ansatz
\begin{equation}\label{3.ans1}
\beta F_{\Gamma, N} = A_\Gamma N + K_\Gamma \log N + C_\Gamma +D_\Gamma/N.
\end{equation}
The results of this fit are presented in Table \ref{t3.3}. At $\Gamma = 4$
this markedly improves the convergence rate, with the final estimate of
$K$ now differing from $1/6$ by only 3 parts in $10^4$. However at
$\Gamma = 6$ the convergence rate is in fact worsened, indicating some
illconditioning when the extra free parameter is introduced. Note also that
the coefficient of $1/N$ in both cases appears to be non-zero, as
distinct from the situation at $\Gamma = 2$ exhibited by the analytic
result (\ref{3.fr2}).

\begin{table}
\caption{\label{t3.3} Fitting the values of $\beta  F_{\Gamma, N}$ with
$N$ as specified, taken from Table \ref{t3.1}, to the ansatz (\ref{3.ans1}).
}

\vspace{.4cm}
{\small
\begin{tabular}{l|l|l|l|l|lr|l|l|l}
$N$ & $A_4$ & $K_4$ & $C_4$ & $D_4$ &  \qquad & 
$A_6$ & $K_6$ & $C_6$ & $D_6$\\ \hline
3,4,5,6 & -2.450743& 0.175200&  0.258672 & 0.049566 & & 
-3.5382 & 0.3594 & 0.0572& 0.4839 \\
4,5,6,7 &  -2.449773 & 0.165568 & 0.2740449 & 0.025973 &&
 -3.5086& 0.0654 & 0.4121 &  0.2363 \\
5,6,7,8 & -2.449905 & 0.167146 & 0.2712323 & 0.031065 && 
 -3.5193 & 0.1932 &  0.1842 & 0.1417 \\
6,7,8,9 & -2.449914 & 0.167268 & 0.2709949 & 0.031065 &&
-3.5180 & 0.1748 &  0.2199 &  0.0779 \\
7 ,8,9,10 &  -2.449896 &  0.166989 &  0.2715743 & 0.029956 &&
 & & & \\
8,9,10,11 &  -2.449892 &  0.166917 & 0.2717321 & 0.029634 \\
& & & \\
\end{tabular}
}
\end{table}

Finally, we sought to estimate from our data an accurate as possible value
of the free energy per particle, $\beta f_\Gamma$ say. 
For this purpose we fitted
the data to the ansatz
\begin{equation}\label{3.ans2}
\beta F_{\Gamma, N} = A_\Gamma N + {1 \over 6} \log N + C_\Gamma +D_\Gamma/N
+ \left \{ \begin{array}{ll} E_\Gamma/N^2, & \Gamma = 4 \\
0, & \Gamma = 6 \end{array} \right.
\end{equation}
thus assuming the universal term in (\ref{3.F1}). Four free parameters
are used at $\Gamma = 4$, while only 3 free parameter are used at 
$\Gamma = 6$, in keeping with observed illconditioning when a fourth
parameter is introduced. Our results are presented in Table \ref{t3.4},
where $\beta f_\Gamma$ is determined by $A_\Gamma$.
We see that there at $\Gamma = 4$ we appear to have convergence to 7
digits with the estimate
\begin{equation}\label{3.e1}
\beta f_4 = -2.449884 \cdots
\end{equation}
while at $\Gamma = 6$ our final estimate is
\begin{equation}\label{3.e2}
\beta f_6 = -3.5175 \cdots
\end{equation}
accurate to 5 digits.

\begin{table}
\caption{\label{t3.4} Fitting the values of $\beta  F_{\Gamma, N}$ with
$N$ as specified, taken from Table \ref{t3.1}, to the ansatz (\ref{3.ans2}).
}

\vspace{.4cm} {\small 
\begin{tabular}{l|l|l|l|lr|l|l|l|l}
$N$ & $A_4$ & $C_4$ & $D_4$ & $E_4$ & \qquad & $A_6$ & $C_6$ & $D_6$ \\
\hline
3,4,5,(6) & -2.4501031 & 0.275576 & 0.012460 & 0.026276  &&
-3.513916& 0.205966& 0.110598 \\ 
4,5,6,(7) &  -2.4498406 & 0.271639 & 0.031880 & 0.005215 && 
-3.518863& 0.250494&0.011648 \\
5,6,7,(8) &  -2.4498809 & 0.272364 & 0.027574 & 0.003235 && 
-3.517146 &  0.231609 & 0.063153\\
6,7,8,(9) &  -2.4498875 & 0.272503 & 0.026605 & 0.005465 && 
-3.517466& 0.235770& 0.049709 \\
7,8,9,(10) & -2.4498842 & 0.272423 & 0.027240 & 0.003788 && 
-3.517540& 0.236870& 0.045600 \\
8,9,10,11 &  -2.4498841 & 0.272420 & 0.027272 & 0.003695 &&
 & & \\
\end{tabular}
}
\end{table}

We note that there is some early literature on estimating $\beta f_4$ and
$\beta f_6$ from exact small $N$ numerical data \cite{JM83}. 
Using only the values of
$\beta F_{N,\Gamma}$ for $N=1,2$ and 3, the quantity
$$
\beta \tilde{f}_\Gamma = \beta f_\Gamma + \Big ( {3 \Gamma \over 8} + 1
\Big ) + {\Gamma \over 4} \log \pi \rho_b - \log \rho_b
$$
was estimated for $\Gamma = 4,6,\dots,10$. In particular, at $\Gamma = 4$ and
$\Gamma = 6$ these estimates of $\beta f_\Gamma$ give
$$
\beta f_4 \approx -2.1585, \qquad \beta f_6 \approx -3.330,
$$
which differ from our estimates (\ref{3.e1}) and (\ref{3.e2}) in the
first decimal place.

\subsection{Free energy -- disk geometry}
For the disk geometry, the prediction (\ref{1.F}) gives a large $N$
expansion of the form
\begin{equation}\label{3.F}
\beta F_\Gamma = AN + BN^{1/2} + {1 \over 12} \log N + C + \cdots
\end{equation}
As in the case of the sphere geometry, this prediction can be verified
analytically using the exact solution for the isotherm $\Gamma = 2$
\cite{AJ81}. The exact solution gives \cite{JMP94}
\begin{equation}\label{3.F1'}
\beta F_2^{\mathrm{hard}} = \beta f_2 N  + \beta 
\gamma_2 N^{1/2} + {1 \over 12} \log N
+ {\mathrm{O}}(1)
\end{equation}
where 
$$
\beta f_2 = {1 \over 2} \log (\rho_b / 2 \pi^2), \quad
\beta \gamma_2 = \sqrt{2}
\int_0^\infty dy \, \log \Big ( {1 \over 2} (1 + {\mathrm{erf}} \, y) \Big ).
$$
Some details of the expansion of $\beta F_2$
 are different for the soft edge
version of the OCP in a disk (recall Section 1). From the exact formula
$$
Z_{N,2}^{\mathrm{soft}} = \pi^N e^{3 N^2 / 4} N^{-N^2/2} (\pi \rho_b )^{-N/2}
G(N+1)
$$
and the asymptotic expansion (\ref{3.barnes}) we see that
\begin{equation}\label{3.F2}
\beta F_2^{\mathrm{soft}} = \beta f N + {1 \over 12} \log N -
\zeta ' (-1) - {1 \over 720 N^2} + {\mathrm{O}} \Big ( {1 \over N^4} \Big ).
\end{equation}
Thus indeed both (\ref{3.F1}) and (\ref{3.F2}) contain the universal term
$(1/12) \log N$, although (\ref{3.F2}) does not contain a surface
tension term (this fact has been noted previously in \cite{DGIL94}).

At $\Gamma = 4$ and $\Gamma = 6$ we obtained exact numerical evaluation of
the partition functions (\ref{3.sa}), (\ref{3.sb}) and (\ref{3.sc}) (and the
modification of (\ref{3.sc}) for the soft disk case) as in the sphere case.
Our results for the corresponding value of $\beta F$ are contained in Table
\ref{t3.5}. To test the prediction (\ref{3.F1}), we sought to fit our
data to the ansatz
\begin{equation}\label{3.F3}
\beta F_{N,\Gamma} = \beta f_\Gamma N + 
B_\Gamma N^{1/2} + K_\Gamma \log N + C_\Gamma + 
 \left \{ \begin{array}{ll} D_\Gamma/N, & {\mathrm{soft \, disk}} \\
0, & {\mathrm{hard \, disk}} \end{array}\right.
\end{equation}
where $\beta f_\Gamma$ is given by (\ref{3.e1}) and (\ref{3.e2}) for
$\Gamma = 4$ and $\Gamma = 6$ respectively, and the choice in
(\ref{3.F3}) is made retrospectively on the criterium of obtaining
better convergence.

Our results are obtained in Table \ref{t3.6}. We see that for the hard
disk at $\Gamma = 4$ our final estimate of $K_4$ differs from 
$1/12$ by only $2 \times 10^{-4}$. At $\Gamma = 6$ we see that more data
would be needed to get a stable sequence, although the final estimates
of $K_6$ are consistent with the expected value of $1/12$.

\begin{table}
\caption{\label{t3.5} Exact decimal expansion of the free energy for the
hard and soft disk at $\Gamma = 4$ and $\Gamma = 6$
}
\vspace{.4cm}
{\small
\begin{tabular}{l|l|l}
$N$ & $F_{N,4}^{\mathrm{hard}}$ & $\beta F_{N,4}^{\mathrm{soft}}$ \\
\hline
3 &  -6.07705853011644579848828232852953 &
-6.38430353764202167882687100789504 \\
4 &  -8.30894530308837749094468707356467 &
 -8.67246771929839719253598118439664 \\
5 &  -10.5685824419856069054748395707000 & 
-10.9817913623032469741300225072724 \\
6 & -12.8480499008173510151377678768908 & 
-13.3060582270200975291371029447052 \\
7 & -15.1423987396644292302500680775824 &
-15.6414978836634761215222474874096 \\
8 & -17.4483520149155330139161065965798 & 
-17.9856458201720068377211714643235 \\
9 & -19.7636864904052121059815096874218 & 
-20.3368227969313363262711724430690 \\
10 & -22.0868149972503557220763154840028 &
-22.6938278975627003536283880871543
\end{tabular}
}
\vspace{.4cm}

\vspace{.4cm}
{\small
\begin{tabular}{l|l|l}
$N$ & $F_{N,6}^{\mathrm{hard}}$ & $\beta F_{N,6}^{\mathrm{soft}}$ \\
\hline
3 &  -9.0582041809587470427592556776938317
& -9.1916690110088058948684895153913657 \\
4 &  -12.306265058620940233015626198772823
&  -12.467150515773535356614120708869630 \\
5 &  -15.583591141405785588643527765993475
& -15.769625685129047660199805300971936 \\
6 & -18.886678348734296934648840469575921
& -19.095091912250933709000748332754870 \\
7 &  -22.209056127812161704085770192533417
& -22.437790137971372572352358156488860 \\ 
8 &  -25.545482070626355796809539664033139
& -25.793196864919170855940747024418097
\end{tabular}
}

\end{table}

\begin{table}
\caption{\label{t3.6} Fit of the ansatz (\ref{3.F3}) to the data of
Table \ref{t3.5}
}

\vspace{.4cm}
{\small
\begin{tabular}{l|l|l|lr|l|l|l|l}
$N$ & $B_4^{\mathrm{hard}}$ & $K_4^{\mathrm{hard}}$ & $C_4^{\mathrm{hard}}$ &
 \quad & $B_4^{\mathrm{soft}}$ & $K_4^{\mathrm{hard}}$ & $C_4^{\mathrm{soft}}$ &
$D_4^{\mathrm{soft}}$ \\ \hline
3,4,5,(6) & 0.749371& 0.059801& -0.091054 & &
0.497409& 0.120202& -0.052807&0.073687 \\ 
4,5,6,(7) & 0.728988& 0.081365& -0.080181 &&
0.509625& 0.099801& -0.040616& 
  0.040317 \\
5,6,7,(8) & 0.723951& 0.087261& -0.078408 &&
0.522124& 0.076905& -0.022675& 
  -0.004884 \\
6,7,8, (9) & 0.726340& 0.084219& -0.078810 &&
0.521397& 0.078345& -0.024029 & 
  -0.001556 \\
7,8,9,(10) & 0.727263& 0.082957& -0.078795 &&
0.518587& 0.084298& -0.030427 & 
  0.014202 \\
8,9,10 & 0.726874& 0.083523& -0.078873 &&
& & & \\
\end{tabular}
}

\vspace{.4cm}
{\small
\begin{tabular}{l|l|l|lr|l|l|l|l}
$N$ & $B_6^{\mathrm{hard}}$ & $K_6^{\mathrm{hard}}$ & $C_6^{\mathrm{hard}}$
&  \qquad & $B_6^{\mathrm{soft}}$ & $K_6^{\mathrm{soft}}$ & $C_6^{\mathrm{soft}}$ 
\\
\hline
3,4,5 & 
1.104506& -0.092158& -0.317518 &&
0.967066& -0.059461& -0.248851 \\
4,5,6 &
0.884919& 0.140146& -0.200388 &&
0.795791& 0.121733& -0.157491 \\
5,6,7 &
0.874984& 0.151776& -0.196890 &&
0.786513& 0.132593& -0.154224 \\
6,7,8 &
0.951461& 0.054407& -0.209757 &&
0.842635& 0.061139& -0.163667 \\
\end{tabular}
}
\end{table}

\subsection{Density and two-point distribution}
\subsection*{Density}
\begin{figure}
\epsfbox{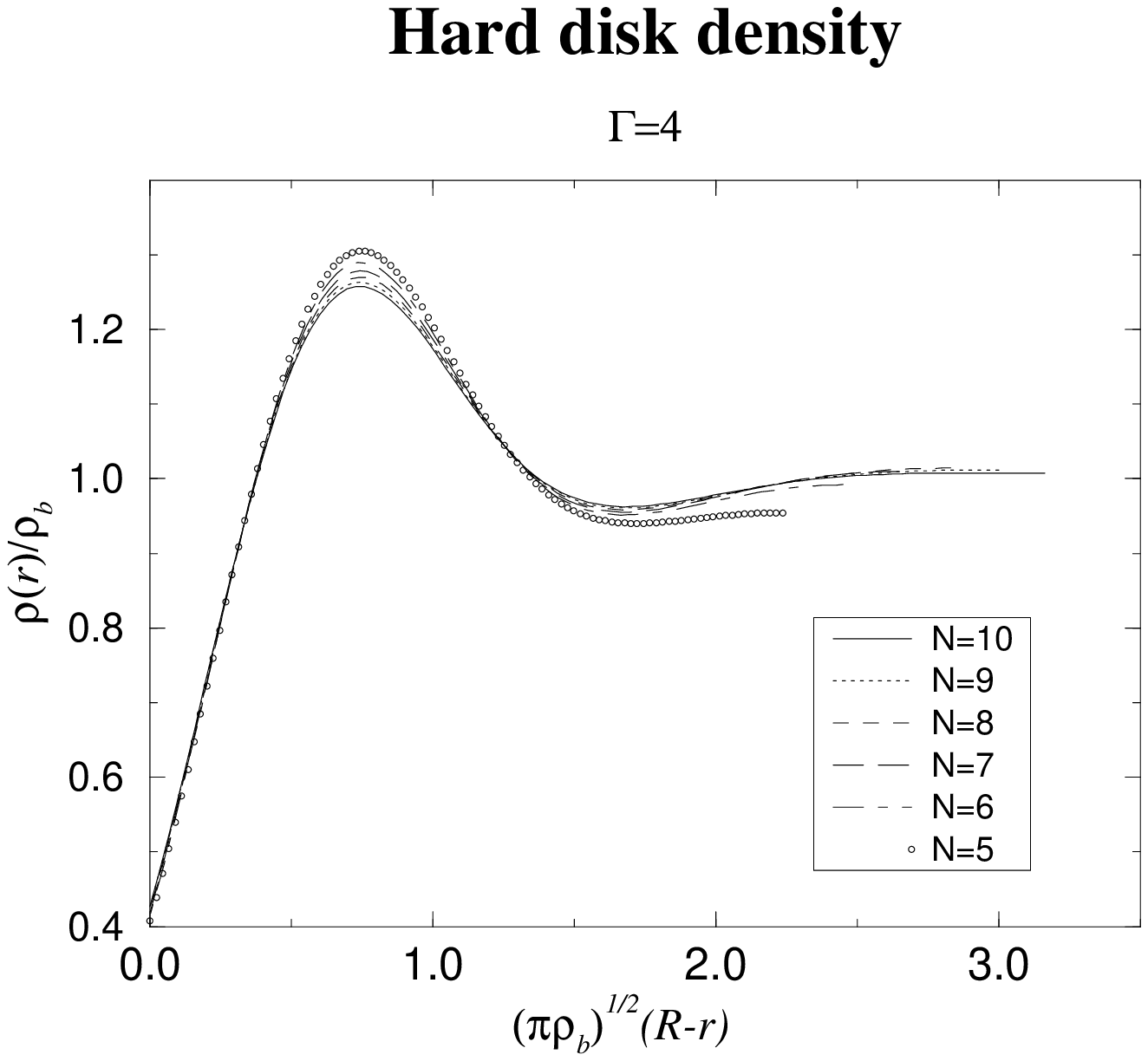}
\caption{\label{f3.1}
Density profile in the hard disk case for several values of $N$ at
$\Gamma=4$. The boundary of the disk is taken as origin.
}
\end{figure}

Consider for definiteness the disk geometry with a hard wall at $\Gamma = 4$.
Using the formula (\ref{2.den}) the density profile can be calculated for
up to 10 particles. One way to present the data is in graphical form
with the boundary of the disk taken as the origin. This is done in
Figure \ref{f3.1}. The plot shows rapid convergence of the profiles near
the boundary.

%
%
%

To investigate the rate of convergence of the whole profile as measured
from the boundary to the thermodynamic value we can investigate the
contact theorem \cite{CFG80}. This expresses the thermodynamic pressure in
terms of the density at contact with the wall, and the potential drop
across the interface (which in turn is proportional to the first moment
of the density profile). Explicitly the contact theorem states
\begin{equation}\label{3.con}
\Big ( 1 - {\Gamma \over 4} \Big ) \rho_b = \rho (0) - 2 \pi \rho_b \Gamma
\int_0^\infty x (\rho (x) - \rho_b ) \, dx
\end{equation}
where we stress again that the density is measured from the boundary.

Much to our initial surprise, the convergence of the r.h.s.~to the l.h.s.~for
the finite $N$ data is very slow. For 10 particles the error is of order
$30 \%$. Further investigation reveals that this is not special to the
coupling $\Gamma = 4$. At $\Gamma = 2$ we have the analytic expression
\cite{Ja81}
$$
\rho_{(1)}(r) = {1 \over 2 \pi}
\sum_{j=1}^N { (R - r)^{2j-2} e^{- \pi (R-r)^2}
\over \int_0^R s^{2j - 1} e^{- \pi s^2} \, ds}, \quad 0 \le r \le R
$$
where $r$ is measured from the boundary
and the background density is taken to equal unity. 
Choosing $N=10$ and substituting
in (\ref{3.con}) again gives an error of order $30\%$. Indeed choosing
$N = 500$ still gives an error of order $3\%$.

In fact the slow convergence of (\ref{3.con}) can be understood analytically
by making use of a sum rule for the OCP applicable for the finite disk
\cite{CFG80}. This sum rule reads
\begin{equation}\label{3.sumr}
\rho_{(1)}(0) - \Big ( 1 - {\Gamma \over 4} \Big ) \rho_b  =
- {\Gamma \rho_b^2 \pi^2 \over N} \int_0^R r^3 \Big (\rho_{(1)}(R - r) - 
\rho_b \Big ) \, dr
\end{equation}
where $\rho_{(1)}(r)$ is measured inward from the boundary. Noting that
charge neutrality requires
$$
\int_0^R r \Big ( \rho_{(1)}(R-r) - \rho_b \Big ) \, dr =
\int_0^R  ( R - r)  \Big ( \rho_{(1)}(r) - \rho_b \Big )  \, dr
$$
we can write
\begin{eqnarray*}\lefteqn{
-  {\Gamma \rho_b^2 \pi^2 \over N} \int_0^R r^3 \Big (\rho_{(1)}(R - r) -
\rho_b \Big ) \, dr} \\
&& = -  {\Gamma \rho_b^2 \pi^2 \over N} \int_0^R (R - r)^3
 \Big (\rho_{(1)}(r) - \rho_b \Big ) \, dr \\
&&  = -  {\Gamma \rho_b^2 \pi^2 \over N} \int_0^R ( - 2r R^2 + 3r^2 R - r^3)
 \Big (\rho_{(1)}(r) - \rho_b \Big ) \, dr \\
&& = 2 \Gamma \rho_b \pi  \int_0^R r \Big (\rho_{(1)}(r) - \rho_b \Big ) \, dr
- {3 \Gamma (\rho_b \pi)^{3/2} \over N^{1/2}}
\int_0^R r^2  \Big (\rho_{(1)}(r) - \rho_b \Big ) \, dr \\ && \quad +
{\Gamma \rho_b^2 \pi^2 \over N} \int_0^R r^3 \Big ( \rho_{(1)}(r) - \rho_b ) \, dr
\end{eqnarray*}
This shows that the finite size corrections to the r.h.s.~of (\ref{3.sumr})
are proportional to $N^{-1/2}$, thus explaining our empirical observation.

\subsection*{Two-point function}
\begin{figure}
\epsfbox{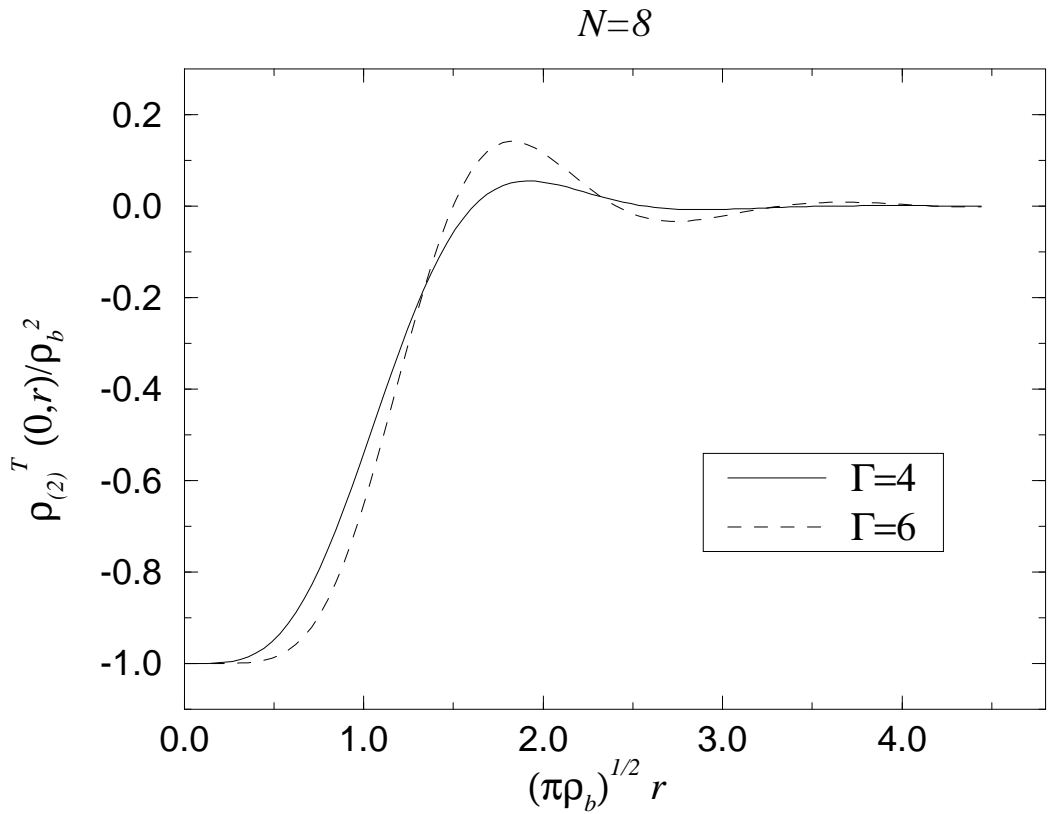}
\caption{\label{f3.2}
Two-point correlation in the sphere case for $N=8$ particles at
$\Gamma=4$ and $\Gamma=6$.
}
\end{figure}

At $\Gamma = 2$ and in the thermodynamic limit the two-particle distribution
function has the exact evaluation \cite{Ja81}
$$
\rho_{(2)}({0}, \vec{r}) = \rho_b^2 \Big ( 1 - e^{-\pi \rho_b |\vec{r}|^2}
\Big ).
$$
This is a monotonic function, with the corresponding truncated distribution
$\rho_{(2)}^T({0}, \vec{r}) := \rho_{(2)}({0}, \vec{r})  - 
\rho_{(1)}({0})\rho_{(1)}(\r)$
exhibiting Gaussian decay to zero. There is evidence, both analytic and
numerical \cite{Ja81,CLWH82} which suggests that for $\Gamma > 2$ the
two-particle distribution exhibits oscillations. At $\Gamma = 4$ this
feature has already been observed in the exact finite $N$ calculation
of $ \rho_{(2)}({0}, \vec{r})$ by Samaj et al.~\cite{SPK94}. Furthermore,
this feature should become more pronounced as $\Gamma$ increases. This is
indeed what we observe when plotting our results for $\Gamma = 4$ and
$\Gamma = 6$ on the same graph (see Figure \ref{f3.2}).

%
%
%

The fact that the 2dOCP is a Coulomb system in its conductive phase implies
that in the bulk the second moment of the truncated distribution obeys
the Stillinger-Lovett sum rule
\begin{equation}\label{3.still}
\int_{{\mathbf{R}}^2} \vec{r}\,{}^2 \rho_{(2)}^T(0,\vec{r}) \, d\vec{r} =
- {2 \over \pi \Gamma}.
\end{equation}
For the hard disk in the finite system we can compute
\begin{equation}\label{3.stillf}
\int_{|\vec{r}| < R} \vec{r}\,{}^2
\rho_{(2)}^T({0}, \vec{r}) \,  d\vec{r}
\end{equation}
and compare it with the universal value given by (\ref{3.still}). At $\Gamma
=4$ and with $N=9$ we find agreement with the universal value to within
$2\%$. In fact, analogous to the integral in (\ref{3.sumr}), the integral
(\ref{3.stillf}) can be evaluated exactly and the terms which differ
from $-2/\pi \Gamma$ read off. In the hard wall case we find
\begin{equation}\label{3.gs}
\int_{|\vec{r}| < R} \r^2
\rho_{(2)}^T(\vec{0}, \vec{r}) \,  d\vec{r}
= - {2 \over \pi \Gamma} \bigg (
\rho_{(1)}(0) / \rho_b + N \rho_{(2)}^T(0,R)/ \rho_b^2 \bigg ),
\end{equation}
while in the soft wall case the same expression results except that the
boundary term $N \rho_{(2)}^T(0,R)/ \rho_b^2$ is no
longer present on the r.h.s., while on the l.h.s.~the integral is over
${\mathbf{R}}^2$.

We see from (\ref{3.gs}) that the deviation in the finite system from the
bulk value (\ref{3.still}) is determined by
$$
- {2 \over \pi \Gamma} \Big ( (\rho_{(1)}(0) - \rho_b)/\rho_b +
N \rho_{(2)}^T(0,R) / \rho_b^2 \Big ),
$$
and thus consists of a bulk and surface contribution.

\section{New sum rules}\label{sec:sum-rules}

In this section we present the derivation of the sum rule~(\ref{3.gs})
and its generalization to multicomponent Coulomb systems. First we
show that the sum rule can be derived within the formalism of
Section~\ref{sec:formalism}, then we present a more general derivation
of the sum rule.

\subsection{The case $\Gamma$ even}

The formalism presented in Section~\ref{sec:formalism} is valid only
if $\Gamma$ is an even integer. Within this formalism we can use the
expressions~(\ref{eq:2.rho2-g4}) and~(\ref{eq:2.rho2-g6}) for the
two-point correlation functions (and its generalizations to higher
$\Gamma$) to compute the second moment
\begin{equation}\label{eq:4.sec-mom}
\int_{\Lambda} \r^2 \rho_{(2)}^T(0,\r)\,d\r\,,
\end{equation}
where $\Lambda$ is a disk of radius $R$ (hard disk) or ${\mathbf{R}}^2$
(soft disk).
For example in the hard disk case with $\Gamma=4$, for each term in the
sum~(\ref{eq:2.rho2-g4}) the integral~(\ref{eq:4.sec-mom}) gives an
incomplete gamma function $\gamma(\mu_j+2,2N)$. Then we use the
recurrence relation
\begin{equation}\label{eq:4.rec}
\gamma(\mu_j+2,2N)=(\mu_j+1)\gamma(\mu_j+1,2N)
-e^{-2N}(2N)^{\mu_j+1}
\end{equation}
to split the expression in two. The first term is proportional to
$\rho_{(1)}(0)$ while the second is proportional to
$\rho_{(2)}(0,R)$. The sum rule~(\ref{3.gs}) follows from that. 

The calculation can be easily generalized to any even $\Gamma$. In the
soft disk case since the incomplete gamma functions are replaced by
complete gamma functions the recurrence relation~(\ref{eq:4.rec}) does
not have a second term on the r.h.s., therefore there is no surface
contribution proportional to $\rho_{(2)}^T(0,R)$ in the sum rule.

\subsection{General case}

In fact a more general derivation of this sum rule, valid for any
value of the coupling constant, can be obtained by
studying the variations of the density as a function of the size of the
disk.

Let us consider the general case of a multicomponent jellium in $\nu$
dimensions confined in a spherical domain $\Lambda$ of radius $R$ and
volume $V=\Omega_\nu R^\nu/\nu$ with
$\Omega_\nu=2\pi^{\nu/2}/\Gamma(\nu/2)$. The system is composed of $s$
different species with charges $(e_\alpha)_{\alpha\in\{1,\dots,s\}}$
and there are $N_\alpha$ particles of the species $\alpha$. Let
$N=\sum_\alpha N_\alpha$ be the total number of particles and let us
define the average density of the species $\alpha$,
$\rho_\alpha=N_\alpha/V$ and the total average density $\rho=N/V$.  As
in the preceding sections $\rho_b$ is the background number density
and let $e_b$ be its charge so that the background charge density is
$e_b\rho_b$. For convenience let us define the ``number of particles of
the background'' by $N_b=\rho_b V$. In general the Coulomb potential
is
\begin{equation}\label{eq:defPhi}
\Phi(\r)=\cases{-\ln r,&if $\nu=2$\cr
\displaystyle\frac{r^{2-\nu}}{\nu-2},&otherwise,}
\end{equation}
and the Coulomb force is
\begin{equation}\label{eq:defF}
\F(\r)=-\nabla\Phi(\r)=\frac{\r}{r^\nu}\,.
\end{equation}
The Hamiltonian of the Coulomb system is
\begin{equation}\label{eq:defU}
U=\frac{1}{2}\sum_{i\neq j}e_{\alpha_i}e_{\alpha_j}\Phi(\r_i-\r_j)
+e_b\rho_b\sum_{i=1}^N e_{\alpha_i}\int_\Lambda d\r \Phi(\r_i-\r)
+\frac{e_b^2\rho_b^2}{2}\int_{\Lambda^2} d\r d\r' \Phi(\r-\r')\,.
\end{equation}
We shall consider the correlation functions in the canonical ensemble
\begin{equation}
\rho_{\alpha_1\dots\alpha_n}^{(n)}(\r_1,\dots,\r_n)=\left<
\sum_{i_1=1}^{N_{\alpha_1}}\cdots
\sum_{i_n=1}^{N_{\alpha_n}}
\delta(\r_1-\r_{\alpha_1,i_1})\cdots
\delta(\r_n-\r_{\alpha_n,i_n})
\right>\,.
\end{equation}
The $\left<\cdots\right>$ is the average in the canonical ensemble and
in the preceding sums if some $\alpha_a=\alpha_b$ we exclude the term
$i_a=i_b$ as usual. 

In three dimensions in order to have a well defined thermodynamic
limit we shall restrict ourselves to the case where all electric
charges $e_\alpha$ have the same sign and the background carries a
opposite neutralizing charge. In two dimensions we can also consider
systems with charges of different signs and eventually without
background ($\rho_b=0$) if the coupling contants $|\beta e_\alpha
e_\gamma| <2$ for all pair of charges $(e_\alpha,\, e_\gamma)$ of
different signs.

\subsubsection{Contact theorem sum rule}

The derivation of the sum rule for the second moment of the two-point
correlation function is similar to that of the contact theorem for a
spherical domain~\cite{CFG80}. Let us first show here the
generalization of this contact theorem for the multicomponent
jellium. We consider the canonical partition function (times $N!$)
\begin{equation}
\Q=\int_{\Lambda^N} d\r^N \exp(-\beta U)\,,
\end{equation}
as a function of the volume $V$. We shall compute the thermodynamical pressure
$p^{(\theta)}=\partial\log\Q/\partial V$ in two different ways. 
The derivative is done at fixed number of
particles and fixed $N_b$.
In
general using the scaling $\r=V^{1/\nu}\tilde{\r}$ we have
\begin{equation}\label{betap0}
\beta p^{(\theta)}=\frac{\partial \ln \Q}{\partial V}=
\rho-\frac{\beta V^N}{\Q}\int_{\tilde{\Lambda}^N} d\tilde{\r}^N
\frac{\partial U(V^{1/\nu}\tilde{\r})}{\partial V}
e^{-\beta U}\,,
\end{equation}
where $\tilde{\Lambda}$ is a sphere of volume 1.

A first way to compute the derivative of $U$ is by using the general
formula
\begin{equation}\label{eq:virial}
\frac{\partial \Phi(V^{1/\nu}\tilde{\r})}{\partial V}=
-\frac{1}{\nu V}\,\r\cdot \F(\r)\,.
\end{equation}
This gives, together with the definition~(\ref{eq:defU}) of $U$,
\begin{eqnarray}\label{eq:betap1}
\beta p^{(\theta)}&=\rho+
\displaystyle\frac{\beta}{\nu V}&
\left[
\int_{\Lambda^2}d\r d\r' \r\cdot \F(\r-\r')\sum_{\alpha,\alpha'}
e_\alpha e_{\alpha'}\rho_{\alpha\alpha'}^{(2)}(\r,\r')
\right.
\nonumber\\
&&
+e_b\rho_b\int_{\Lambda^2}d\r d\r' (\r-\r')\cdot \F(\r-\r')
\sum_\alpha e_\alpha \rho_\alpha^{(1)}(\r')
\nonumber\\
&&
\left.
+\frac{1}{2}e_b^2\rho_b^2 \int_{\Lambda^2} d\r d\r'
(\r-\r')\cdot \F(\r-\r')
\right]\,.
\end{eqnarray}
We can transform the preceding expression by using the first equation
of the BGY hierarchy
\begin{equation}\label{eq:BGY1}
k_B T\nabla \rho_\alpha(\r)=e_\alpha \rho_b e_b\int_\Lambda d\r'\,
\F(\r-\r')\rho_\alpha^{(1)}(\r)+ \int_\Lambda d\r' 
\sum_{\alpha'} e_\alpha e_{\alpha'} \F(\r-\r')
\rho_{\alpha \alpha'}^{(2)} (\r,\r)\,.
\end{equation}
The r.h.s of~(\ref{eq:BGY1}) appears in the first and second lines
of~(\ref{eq:betap1}). Replacing it by the l.h.s of~(\ref{eq:BGY1}) we
find
\begin{eqnarray}
\beta p^{(\theta)}&=\rho+
\displaystyle\frac{\beta}{\nu V}&
\left[
k_B T \int_{\Lambda} d\r \sum_\alpha \r\cdot\nabla\rho_\alpha^{(1)}(\r)
\right.
\nonumber\\
&&
-e_b \rho_b \int_{\Lambda^2}d\r d\r' \r'\cdot\F(\r-\r') 
\sum_\alpha e_\alpha \rho_\alpha^{(1)}(\r)
\nonumber\\
&&
\left. +\frac{1}{2} \rho_b^2 e_b^2 \int_{\Lambda^2} d\r d\r'\, (\r-\r')\cdot
\F(\r-\r') 
\right]\,.
\end{eqnarray}
The first term of the r.h.s of the preceding equation can be computed
by integration by parts while the others can be computed using the
definition~(\ref{eq:defF}) of the Coulomb force $\F$ and Newton's
theorem. This yields the following expression for the thermodynamical
pressure
\begin{eqnarray}\label{eq:betapfin1}
\beta p^{(\theta)}&=&\sum_\alpha \rho_\alpha^{(1)}(R)
+\beta e_b \rho_b R^{2-\nu}\left(\sum_\alpha e_\alpha 
\frac{N_\alpha}{2}
+e_b\frac{N_b}{\nu+2}\right)
\nonumber\\
&&
-\frac{\beta\rho_b e_b}{2R^\nu}
\int_\Lambda d\r\, r^2 \sum_\alpha e_\alpha \rho_\alpha^{(1)}(\r)
\,.
\end{eqnarray}

The other way to compute the thermodynamical pressure is to use the
actual scaling properties of the Coulomb potential $\Phi$,
\begin{equation}\label{eq:scalPhi}
\frac{\partial\Phi(V^{1/\nu}\tilde{\r})}{\partial V}
=\cases{\displaystyle-\frac{1}{2V},&if $\nu=2$\cr
\displaystyle\frac{2-\nu}{\nu V}\,\Phi(\r),&otherwise.
}
\end{equation}
Substituting this expression in~(\ref{betap0}) gives
\begin{equation}\label{eq:betapfin2}
\beta p^{(\theta)}=\rho+
{\beta \delta_{\nu,2} \over 4} \left(\frac{ Q^2}{V}
-\sum_\alpha  e_\alpha^2 \rho_\alpha
\right)
+ \frac{\nu-2}{\nu V}\beta \left<U\right>\,,
\end{equation}
where $Q=\sum_\alpha e_\alpha N_\alpha+ e_b N_b$ is the total charge of
the system.

Equating the two expressions~(\ref{eq:betapfin1})
and~(\ref{eq:betapfin2}) of the thermodynamic pressure we find the
generalization of the contact theorem
\begin{eqnarray}\label{eq:contact}
\lefteqn{
\sum_\alpha \rho_\alpha^{(1)}(R)+\frac{\beta e_b \rho_b}{2R^{\nu-2}}Q
-\frac{\beta e_b \rho_b}{2R^\nu}\int_\Lambda d\r\, r^2 q(\r)
}
\nonumber\\
&&= \rho  + \delta_{\nu,2} \frac{\beta}{4} 
\left(\frac{Q^2}{V}-\sum_\alpha e_\alpha^2 \rho_\alpha\right)
+\frac{\nu-2}{\nu V} \beta \left<U\right>
\end{eqnarray}
where $q(\r)=\sum_\alpha e_\alpha \rho_\alpha^{(1)}(\r)+e_b \rho_b$ is
the local charge density.

\subsubsection{Density-charge correlation second moment sum rule}

Similar calculations lead to the second moment sum rule for the
density-charge truncated correlation function $\sum_\beta
e_\beta\rho_{\alpha\beta}^{(2)T}(0,\r)$. Here we consider the quantity
\begin{equation}
\Q_\alpha = \int_{\Lambda^N} d\r^N e^{-\beta U}
\sum_{i=1}^{N_\alpha}
\delta(\r_{i,\alpha})
\,,
\end{equation}
as a function of the volume $V$. Note that the density of the species
$\alpha$ at the center of the spherical domain is
$\rho_\alpha^{(1)}(0)=\Q_\alpha/\Q$. Like in the preceding section we
want to compute by two different ways the quantity $\Q^{-1}\partial
\Q_\alpha/\partial V$. Using the same scaling argument as before we
have
\begin{equation}\label{eq:dQalpha0}
\frac{1}{\Q}\frac{\partial\Q_\alpha}{\partial V}=
\frac{N-1}{V}\rho_\alpha^{(1)}(0)
-\beta \frac{V^N}{\Q}
\int_{\tilde{\Lambda}^N} d\tilde{r}^N
\sum_{i=1}^{N_\alpha} \delta(V^{1/\nu}\tilde{\r}_{i,\alpha})
\,\frac{\partial U(V^{1/\nu}\tilde{\r})}{\partial V}
\,e^{-\beta U}
\,.
\end{equation}
Using eq.~(\ref{eq:virial}) and the definition~(\ref{eq:defU}) of the
Hamiltonian $U$ we find
\begin{eqnarray}\label{eq:qalpha1}
\frac{1}{\Q}\frac{\partial\Q_\alpha}{\partial V}&=&
\frac{N-1}{V}\rho_\alpha^{(1)}(0)
\nonumber\\
&&+\frac{\beta}{\nu V}
\left[
\int_{\Lambda^2}d\r d\r'\, \r\cdot\F(\r-\r')
\sum_{\beta,\gamma} e_\beta e_\gamma \rho_{\alpha\beta\gamma}^{(3)}
(0,\r,\r')
\right.
\nonumber\\
&&+\int_\Lambda \r\cdot\F(\r) e_\alpha \sum_\beta e_\beta
\rho_{\alpha\beta}^{(2)}(0,\r)
\nonumber\\
&&+e_b \rho_b \int_{\Lambda^2}d\r d\r'\sum_\beta e_\beta
\rho_{\alpha\beta}^{(2)}(0,\r)\, \r\cdot\F(\r-\r')
\nonumber\\
&&-e_b \rho_b \int_{\Lambda^2} d\r d\r' \r'\cdot\F(\r-\r')
\sum_\beta e_\beta \rho_{\alpha\beta}^{(2)}(0,\r)
\nonumber\\
&&+e_b \rho_b\int_\Lambda d\r\,\r\cdot\F(\r) e_\alpha
\rho_\alpha^{(1)}(0)
\nonumber\\
&&
\left.
+\frac{1}{2} e_b^2\rho_b^2\int_{\Lambda^2} d\r d\r'\,
(\r-\r')\cdot\F(\r-\r') \rho_\alpha^{(1)}(0)
\right]
\end{eqnarray}
Using the second BGY equation
\begin{eqnarray}\label{eq:BGY2}
k_B T\nabla_\r\rho_{\alpha\beta}^{(2)}(0,\r)&=&e_\beta e_b \rho_b
\int_\Lambda d\r'\,\F(\r-\r')\rho_{\alpha\beta}^{(2)}(0,\r)
\nonumber\\
&&+e_\beta e_\alpha \F(\r) \rho_{\alpha\beta}^{(2)}(0,\r)
\nonumber\\
&&+\int_\Lambda d\r'\,\F(\r-\r')\sum_\gamma e_\beta e_\gamma
\rho_{\alpha\beta\gamma}^{(3)}(0,\r,\r')
\,,
\end{eqnarray}
and then integration by parts
\begin{equation}
\sum_\beta \int_\Lambda \r\cdot\nabla_\r\rho_{\alpha\beta}^{(2)}
(0,\r)=\nu V\sum_\beta \rho_{\alpha\beta}^{(2)}(0,R)
-\nu(N-1)\rho_\alpha^{(1)}(0)
\,,
\end{equation}
we can arrange expression~(\ref{eq:qalpha1}) to find, after computing
explicitly the integrals involving $\F$ using Newton's theorem,
\begin{eqnarray}\label{eq:dQalpha1}
\frac{1}{\Q}\frac{\partial\Q_\alpha}{\partial V}&=&
\sum_\beta \rho_{\alpha\beta}^{(2)}(0,R)
-\frac{\beta e_b\rho_b}{2R^\nu}
\int_\Lambda d\r\,r^2\sum_\beta e_\beta \rho_{\alpha\beta}^{(2)}(0,\r)
\nonumber\\
&&+\frac{\beta e_b\rho_b}{2R^{\nu-2}}\rho_\alpha^{(1)}(0)
\left[\sum_\beta e_\beta N_\beta+\frac{e_b N_b}{\nu+2}\right]
\,.
\end{eqnarray}

The second way for computing $\Q^{-1}\partial\Q_\alpha/\partial V$ is
by using directly equation~(\ref{eq:scalPhi}) into
equation~(\ref{eq:dQalpha0}). This gives,
\begin{eqnarray}\label{eq:dQalpha2}
\frac{1}{\Q}\frac{\partial\Q_\alpha}{\partial V}&=&
\frac{N-1}{V}\rho_\alpha^{(1)}(0)+\frac{\beta}{4}\delta_{\nu,2}
\left[\frac{Q^2}{V}-\sum_\beta e_\beta^2
N_\beta\right]\rho_\alpha^{(1)}(0)
\nonumber\\
&&
+\frac{\nu-2}{\nu V}\beta
\left<
U\hat{\rho}_\alpha^{(1)}(0)
\right>
\,,
\end{eqnarray}
where
$\hat{\rho}_\alpha^{(1)}(0)=\sum_{i=1}^{N_\alpha}\delta(\r_{i,\alpha})$
is the microscopic density of $\alpha$-particles at the center of the
domain $\Lambda$.

Comparing the two expressions~(\ref{eq:dQalpha1})
and~(\ref{eq:dQalpha2}) of $\Q^{-1}\partial\Q_\alpha/\partial V$ gives
a sum rule for the second moment of the density of $\alpha$
particles-electric charge correlation function. The sum rule takes a
nice form by considering the truncated correlation function and making
use of the contact sum rule~(\ref{eq:contact}),
\begin{eqnarray}
\label{eq:thenewsumrule}
\frac{\beta e_b \rho_b\Omega_\nu}{2\nu} \int_{\Lambda} d\r\, r^2 \sum_\beta
e_\beta \rho_{\alpha\beta}^{(2)T}(0,\r) 
&=&
\rho_\alpha^{(1)}(0) +\frac{\Omega_\nu}{\nu} R^\nu
\sum_\beta \rho_{\alpha\beta}^{(2)T}(0,R)
\nonumber\\ 
&&+\frac{2-\nu}{\nu}\,
\beta\left<U\hat{\rho}_{\alpha}^{(1)}(0)\right>^T \,.
\end{eqnarray}

In the case of the two-dimensional OCP ($\nu=2$, $s=1$ and $e_b=-e_1$)
this is exactly the sum rule~(\ref{3.gs}) announced in the preceding
section
$$
\int_{|\vec{r}| < R} \r^2
\rho_{(2)}^T(\vec{0}, \vec{r}) \,  d\vec{r}
= - {2 \over \pi \Gamma} \bigg (
\rho_{(1)}(0) / \rho_b + N \rho_{(2)}^T(0,R)/ \rho_b^2 \bigg )
\,.
\eqno\hbox{(\ref{3.gs})}
$$

The sum rule~(\ref{eq:thenewsumrule}) is in fact a series of $s$ sum
rules for the density-charge correlation function $\sum_\beta e_\beta
\rho^{(2)T}_{\alpha\beta}(0,\r)$ for each species $\alpha$. By taking
the sum of these sum rules with the factors $e_\alpha$, we find a sum rule
for the charge-charge truncated correlation function
$S(0,\r)=\sum_{\alpha,\beta} e_\alpha e_\beta
\rho^{(2)T}_{\alpha\beta}(0,\r)$,
\begin{eqnarray}
\label{eq:charge-sumrule}
\frac{\beta e_b \rho_b\Omega_\nu}{2\nu} 
\int_{\Lambda} d\r\, r^2 S(0,\r) 
&=&
\sum_{\alpha}e_\alpha\rho_\alpha^{(1)}(0) 
+\frac{\Omega_\nu}{\nu} R^\nu
\sum_{\alpha,\beta} e_\alpha \rho_{\alpha\beta}^{(2)T}(0,R)
\nonumber\\ 
&&+\frac{2-\nu}{\nu}\,
\beta\left<U\sum_\alpha e_\alpha 
\hat{\rho}_{\alpha}^{(1)}(0)\right>^T \,.
\end{eqnarray}

\subsubsection{Thermodynamic limit of the sum rules}
\subsubsection*{Canonical ensemble}

In order to study the relationship between sum
rules~(\ref{eq:thenewsumrule}) and~(\ref{eq:charge-sumrule}) and the
Stillinger--Lovett sum rule, we need to know the behavior of the
correlation functions as they approach the thermodynamic limit. This
behavior is different depending on the ensemble used. In this
section we continue to work in the canonical ensemble.

In general we shall suppose that in the thermodynamic limit the system
is in a fluid and conducting phase. In
this case the density becomes uniform in the thermodynamic limit
$\rho_\alpha^{(1)}(0)\to \rho_\alpha$ and 
\begin{equation}\label{eq:Urho}
\left<\hat{\rho}_\alpha^{(1)}(0)U\right>^T
\to\left<\rho_\alpha U\right>^T=0\,,
\end{equation}
because in the canonical ensemble the density does not fluctuate.

Let us first consider the case of a multicomponent Coulomb system
without background (in two dimensions with small Coulomb couplings).
In that case equation~(\ref{eq:thenewsumrule})
becomes
\begin{equation}\label{eq:can-rhob-0-fini}
\rho_\alpha^{(1)}(0) +\frac{\Omega_\nu}{\nu} R^\nu
\sum_\beta \rho_{\alpha\beta}^{(2)T}(0,R)
=0\,.
\end{equation}
This
equation~(\ref{eq:can-rhob-0-fini}) give us the behavior of the correlation
functions as they approach the thermodynamic limit
\begin{equation}\label{eq:tails}
\sum_{\gamma} \rho_{\alpha\gamma}^{(2)T} (0,R)
\sim -\rho\rho_\alpha/N
\,.
\end{equation}
This is a generalization of an already known result concerning the
existence of $1/N$ tails for the correlation functions of one
component fluids with short range forces~\cite{LP61}. However, for a neutral
system taking the sum of equations~(\ref{eq:tails}) with the
coefficients $e_\alpha$ show that the charge-total density correlation
does not have $1/N$ tails,
\begin{equation}
R^\nu\sum_{\alpha\gamma} e_\alpha \rho_{\alpha\gamma}^{(2)T} (0,R)
\to 0
\,,
\end{equation}

It is likely that a similar behavior exists in the general case
($\rho_b\neq0$, $\nu=2,3$), so it would be difficult to derive
from~(\ref{eq:thenewsumrule}) partial sum rules for the density-charge
correlations in the thermodynamic limit because with the $1/N$ tails,
one cannot commute the thermodynamic limit with the integration over
the space. However, one can conjecture that although the
density-density correlations have $1/N$ tails, in the conductive phase
the total density-charge correlations do not have these tails as it is
in the case when $\rho_b=0$. If this is true, and assuming that the
convergence of the charge-charge correlation function is uniform (in
order to commute the thermodynamic limit with the integration over the
space), one can recover the Stillinger--Lovett sum rule from the sum
rule~(\ref{eq:charge-sumrule}) for finite systems,
\begin{equation}\label{eq:SL}
\frac{\beta\Omega_\nu}{2\nu}
\int_{{\mathbf{R}}^\nu} d\r r^2 S(0,\r)=-1\,.
\end{equation}
The fact that we recover the Stillinger--Lovett sum rule is of course
not a proof of our conjecture, but at least it show that our
conjecture is not in contradiction with well known results.

\subsubsection*{Grand canonical ensemble}

For systems with short range forces the correlations functions do not
have $1/N$ tails in the grand canonical ensemble as they approach the
thermodynamic limit~\cite{LP61}. We will show that this is also the
case for two-dimensional Coulomb systems with small couplings when
there is no charged background and assuming this is also the case in
general for a multicomponent jellium we will discuss the thermodynamic
limit of the partial sum rules.

The partial sum rules~(\ref{eq:thenewsumrule}) obtained before are
different in the grand canonical ensemble. The grand
canonical ensemble is parametrized by the background density $\rho_b$
and $s-1$ fugacities $\{z_\gamma\}$ used to fix $s-1$ average densities
$\rho_\gamma$, the remaining density fixed by electroneutrality.  The
grand canonical version of the sum rules~(\ref{eq:thenewsumrule}) can
be obtained in a straightforward manner by adapting the calculations of
the last section. However special care should be taken because of the
fluctuation of the average densities in the grand canonical
ensemble. These fluctuations add some extra terms to sum
rule~(\ref{eq:thenewsumrule}),
\begin{eqnarray}
\label{eq:sumrule-gc}
\frac{\beta e_b \rho_b\Omega_\nu}{2\nu} \int_{\Lambda} d\r\, r^2 \sum_\beta
e_\beta \rho_{\alpha\beta}^{(2)T}(0,\r) 
&=&
\left<\hat\rho_\alpha^{(1)}(0)\right> +\frac{\Omega_\nu}{\nu} R^\nu
\sum_\beta \rho_{\alpha\beta}^{(2)T}(0,R)
\nonumber\\ 
&&-\left<N\hat\rho^{(1)}_\alpha(0)\right>^T
+\frac{2-\nu}{\nu}\,
\beta\left<U\hat{\rho}_{\alpha}^{(1)}(0)\right>^T 
\nonumber\\
&&+\delta_{\nu,2}
\left<\sum_\gamma \frac{\beta e_\gamma^2}{4} N_\gamma
\hat\rho_\alpha^{(1)}(0)\right>^T
\,.
\end{eqnarray}

To proceed with the discussion of the thermodynamic limit of this sum
rule, we need to use a relation that will allow us to simplify the
terms on the r.h.s.~of equation~(\ref{eq:sumrule-gc}) in the
thermodynamic limit. This relation reads for $\nu=2$ or 3,
\begin{equation}\label{eq:scaling-relation}
\rho_\alpha
-\left<N\rho_\alpha\right>^T
+\frac{2-\nu}{\nu}\,
\beta\left<U\rho_{\alpha}\right>^T 
+\delta_{\nu,2}
\left<\sum_\gamma \frac{\beta e_\gamma^2}{4} N_\gamma
\rho_\alpha\right>^T
=
\rho_b\frac{\partial\rho_\alpha}{\partial\rho_b}
\,.
\end{equation}
This relation is a consequence of the scaling properties of the
Coulomb potential. To prove it, let us consider the thermodynamic
grand canonical pressure
\begin{equation}
\beta \tilde p(\beta,\{z_\gamma\},\rho_b)
=\lim_{V\to\infty} V^{-1} \ln \Xi(\beta,\{z_\gamma\},\rho_b,V)
\,,
\end{equation}
where $\Xi$ is the grand canonical partition function. Using the
scaling properties of the Coulomb potential we have for $\nu=3$
\begin{equation}
\beta \tilde p(\beta,\{z_\alpha\},\rho_b)=\lambda^4\beta
\tilde p(\lambda\beta,\{\lambda^{-3/2}z_\alpha\},\lambda^{-3}\rho_b)
\,,
\end{equation}
and for $\nu=2$,
\begin{equation}
\beta \tilde p(\beta,\{z_\alpha\},\rho_b)=
\beta \tilde p(\beta,\{\lambda^{-2(1-(\beta e_\alpha^2/4))}z_\alpha\},
\lambda^{-2}\rho_b)
\,,
\end{equation}
for any positive number $\lambda$. Taking the derivative of these
relations with respect to $\lambda$, then putting $\lambda=1$ and using
the usual thermodynamic relations yields for $\nu=3$,
\begin{equation}\label{eq:p-nu3}
\tilde p=\frac{1}{3}\left<H\right> + 
\frac{1}{2\beta} \rho - \rho_b\frac{\partial\tilde p}{\partial\rho_b}
=\frac{1}{3}\left<U\right>+ \frac{1}{\beta} \rho 
-\rho_b\frac{\partial\tilde p}{\partial\rho_b} 
\,,
\end{equation}
and for $\nu=2$,
\begin{equation}\label{eq:p-nu2}
\beta \tilde p=\sum_\alpha \left(1-\frac{\beta e_\alpha^2}{4}\right)
\rho_\alpha
+\beta\rho_b
\frac{\partial\tilde p}{\partial\rho_b}\,.
\end{equation}
where $\left<H\right>$ is the total internal energy (including the
kinetic term). The announced relation~(\ref{eq:scaling-relation})
follows from taking the derivative of~(\ref{eq:p-nu3})
and~(\ref{eq:p-nu2}) with respect to the fugacities.

As before let us consider first the case $\rho_b=0$ (in two dimensions
for systems with small couplings). Then
equation~(\ref{eq:sumrule-gc}) together with
equation~(\ref{eq:scaling-relation}) shows that the grand canonical
total density-partial density correlation function does not exhibit
any $1/N$ tails,
\begin{equation}\label{eq:notails}
R^\nu \sum_{\alpha\gamma} \rho^{(2)T}_{\alpha\gamma}(0,R) \to0\,.
\end{equation}
Now if we suppose that in the general case ($\rho_b\neq0$, $\nu=2,3$) this
property still holds and that the density-charge correlation functions
converge uniformly we recover the partial sum rules
\begin{equation}\label{eq:sumrule-svw}
\frac{\beta e_b\Omega_\nu}{2\nu}
\int_{{\mathbf{R}}^\nu} d\r r^2
\sum_\gamma e_\gamma\rho_{\alpha\gamma}^{(2)T} (0,\r)
=
\frac{\partial\rho_\alpha}{\partial\rho_b}
\,,
\end{equation}
that have been previously derived by Suttorp and
van~Wonderen~\cite{SvW87} in the three dimensional case. These
equations also hold for two-dimensional systems.  One can recover the
Stillinger--Lovett sum rule~(\ref{eq:SL}) by taking the sum of these
equations~(\ref{eq:sumrule-svw}) with the factors $e_\alpha$ and using
electroneutrality. Notice that the condition~(\ref{eq:notails}) on the
thermodynamic limit of the two-point correlation function when one of
the points is in the boundary is different from the usual condition
needed to prove the Stillinger--Lovett~\cite{MG83} that the
correlation function of the infinite system should decay faster than
$1/r^{\nu+2}$.

Notwithstanding the relation of the sum rules~(\ref{eq:thenewsumrule})
and~(\ref{eq:sumrule-gc}) with the Stillinger--Lovett sum
rule~(\ref{eq:SL}), let us stress that for finite systems these
sum rules are not screening sum rules like the Stillinger--Lovett sum
rule since for finite systems the screening of external charges does
not exists (because since the total electric charge is conserved, the
excess of charge can not leak out to infinity like it does in infinite
systems). From the derivation presented in the previous section it is
clear that the new sum rules should be seen more as a second order
contact theorem rather than a screening sum rule. Futhermore when
there is no background ($\rho_b=0$) the relation with
Stillinger--Lovett sum rule disappears because the term containing the
second moment of the density-charge correlation vanishes.

\section{Summary and conclusion}

Expanding the power of the Vandermonde determinant that appears in the
Boltzmann factor of the 2dOCP in terms of simple orthogonal
polynomials we have been able to develop exact numerical solutions for
values of the coupling constant $\Gamma=4$ and $\Gamma=6$ for finite
systems up to 11 and 9 particles respectively for different kinds of
geometry (sphere, soft and hard wall disk). With these solutions we
have been able to test the prediction~\cite{JMP94} of universal
logarithmic finite size corrections to the free energy~(\ref{1.F}).
Studying the correlation functions has lead us to find a new sum
rule~(\ref{3.gs}) similar to the Stillinger--Lovett sum rule for
finite systems. This sum rule can be derived within the formalism of
section~\ref{sec:formalism}, but can also be generalized to higher
dimension and multicomponent jellium systems
(eq.~(\ref{eq:thenewsumrule})).

Further applications of the formalism presented here are the study of
surface correlations which are expected to have a universal behavior
at large distances~\cite{Jan95}. Also the formal expressions of the
correlations functions~(\ref{eq:2.rho2-g4}) and~(\ref{eq:2.rho2-g6})
could eventually be used to find higher order sum rules or other
general properties.

\section*{Acknowledgements}

G.~T.~acknowledges the financial support from the Australian Research
Council and would like to thank the Department of Mathematics and
Statistics of the University of Melbourne for its hospitality.
Also, we are particularly grateful to J.-Y.~Thibon and
B.G.~Wybourne for supplying us with their data from \cite{STW94}.
We thank B.~Jancovici for a useful remark on
section~\ref{sec:sum-rules}.

\end{document}